\documentstyle[aps,twocolumn]{revtex}

\begin{document} 
\draft
\title{Phenomenological Evidence for Gluon Depletion
in $pA$ Collisions}
\author{Rudolph C. Hwa}
\address{Institute of Theoretical Science and Department of Physics\\
University of Oregon, Eugene, OR 97403-5203, USA}
\author{J\'{a}n Pi\v{s}\'{u}t and Neva Pi\v{s}\'{u}tov\'{a}}
\address{Department of Physics, Comenius University, SK-84215, Bratislava,
Slovakia}
\date{\today}
\maketitle
 
\begin{abstract}
The data of $J/\psi$ suppression at large $x_F$ in $pA$ collisions are used to infer
the existence of gluon depletion as the projectile proton traverses the nucleus. The
modification of the gluon distribution is studied by use of a convolution equation whose
non-perturbative splitting function is determined phenomenologically. The depletion
factor at $x_1=0.8$ is found to be about 25\% at $A=100$.
\end{abstract}

It is conventional in the study of $J/\psi$ production in heavy-ion collisions that the 
gluon distribution before the hard subprocess of $c\bar c$ production is assumed
to be the same as in a free nucleon \cite{ks,wj,v}.  The unconventional view that the
gluon distribution  can be modified in the nuclear medium due to depletion was
suggested in \cite{hpp}.  In this paper we focus on
$p$-$A$ collisions and show that the data \cite{l} on $\alpha (x_F)$ can be used to
infer that gluon depletion in the projectile proton is not negligible.  
 
Charmonium absorption in $pA$ collisions has been studied in \cite{hhk,a} without
finding any satisfactory explanation for the $x_F$ dependence of $\alpha(x_F)$.  In
\cite{v2} the effect of energy loss of partons is considered, but that is only one
aspect of gluon depletion.  Here we pay
particular attention to the evolution of the gluon distribution of the projectile as it
traverses the nucleus.  The approximate absence of dilepton suppression and the
consequent implication that the quark distribution is nearly unaltered by the nuclear
medium lead some to expect that the gluon distribution would be unaltered also. 
However, such a view is based on the validity of DGLAP evolution of the parton
distribution functions
\cite{kr}.  We adopt the reasonable alternative view that the evolution in a nucleus is
different from that of pQCD at high $Q^2$; indeed, we shall let the data guide us in
determining the proper dynamics of the low-$Q^2$ non-perturbative process.

The Fermilab E866 experiment measured the $J/\psi$
  suppression in $p$-$A$ collisions at 800 GeV/c with a wide coverage of
$x_F$ \cite{l}.  The result is given in terms of $\alpha(x_F)$, which is defined
by the formula
\begin{eqnarray} R(x_F, A) = \sigma _A(x_F)/A\,\sigma_N(x_F) =
A^{\alpha(x_F)-1}\quad , 
\label{1}
\end{eqnarray} 
where $\sigma_{N, A}$ is the cross section for $J/\psi$ production by a proton  on a
nucleon $(N)$ or on a nucleus $(A)$.  In \cite{l} a parametrization of
$\alpha(x_F)$ for $J/\psi$  production is given:
\begin{eqnarray}
\alpha(x_F) = 0.952 (1 + 0.023 x_F - 0.397 x_F^2)
\label{2}
\end{eqnarray} for $- 0.1 < x_F < 0.9$.  It is our aim here to explore the implication of Eq.
(\ref{2}) on the evolution of the gluon distribution.

Since the semihard subprocess of $g + g \rightarrow c + \bar c$ is common for $p$-$N$ and
$p$-$A$ collisions, they cancel in the ratio
$R(x_F, A)$ so the $x_F$ dependence can come from three sources:  (a) the ratio
of the gluon distribution in the projectile passing through a nucleus to that in a free proton,
$G(x_F, A)$, (b) nuclear shadowing of gluons in the target, $N(x_F,
A)$, and (c) hadronic absorption of the $c
\bar c$ states after the semihard subprocess, $H(x_F, A)$.  Putting them
together, we have
\begin{eqnarray} R(x_F, A) =  G(x_F, A) N(x_F, A)
H(x_F, A) \quad. 
\label{3}
\end{eqnarray}
$G(x_F, A)$ and $N(x_F, A)$ are ignored in
\cite{hhk,a}.  Since $x_F < 0.25$ in \cite{hhk}, there is not much dependence on $x_F$ to
be ascribed to $H(x_F,A)$, but in
\cite{a}, where the full range of $x_F$ is considered, $H(x_F,A)$ is forced to carry
the entire $x_F$-dependence by a fitting procedure, resulting in an unreasonably short
octet lifetime.  Our approach by including $G(x_F, A)$ and $N(x_F, A)$
in Eq.(\ref{3}) is therefore complementary to the work of \cite{hhk,a}.

The nuclear shadowing problem has been studied in detail by Eskola {\it et al.}
\cite{eks,eksa}, using the deep inelastic scattering data of nuclear targets at high $Q^2$.  On
the basis of DGLAP evolution they can determine the parton distributions at any $Q^2 >
2.25$ GeV$^2$.  The results are given in terms of numerical parametrizations (called
EKS98 \cite{eksa}) of the ratio $N^A_i (x, Q^2) = f_{i/A}(x, Q^2)/ f_i
(x, Q^2)$, where $f_i$ is the parton distribution of flavor $i$ in the free proton
and $f_{i/A}$ is that in a proton of a nucleus $A$.  We shall be interested in  the ratio
for the gluon distributions only at $Q^2 = 10\ {\rm GeV}^2$, corresponding to $c \bar c$
production, and denote it by $N(x, A)$.  From the numerical output of EKS98 we find that a
simple formula can provide a good fit to within 2\% error in the range $40 < A < 240$ and
$0.01 <x<0.12$; it is 
\begin{eqnarray} N(x, A) = A^{\beta(x)},
\label{4}
\end{eqnarray} where
\begin{eqnarray}
\beta (\xi (x) ) = \xi\ (0.0284 + 0.0008\,\xi - 0.0041\,\xi^2),
\label{5}
\end{eqnarray} 
with $\xi = 3.912 + {\rm ln} x$.  Thus the $A$ dependence is minimal at
$\xi = 0$, corresponding to $x = 0.02$.

The variable $x$ in Eq.(\ref{4}) is the gluon momentum fraction in a nucleon in the
nucleus, usually referred to as $x_2$.  Both $x_F$ in Eq.(\ref{1}) and $x_2$ in Eq.(\ref{4})
are to be converted to the
$x_1$ variable for the projectile nucleon, using 
\begin{eqnarray} x_F = x_1 - x_2,  \qquad\qquad x_1 x_2 =\tau\equiv M^2_{J / \psi }/s ,
\label{6}
\end{eqnarray}
 so that a part of Eq.\ (\ref{3}) can be rewritten as 
\begin{eqnarray} R(x_F, A)/N(x_2, A) = A^{\alpha ( x_F
(x_1))-\beta (x_2(x_1)) - 1}.
\label{7}
\end{eqnarray} 
In our approach we treat $H(x_F, A)$ as having negligible
dependence on $x_F$ for all $x_F$.  Attempts \cite{hhk,a} to find that
dependence have failed and led to the suggestion of the existence of an unaccounted
mechanism responsible for the enhanced suppression in $R(x_F, A)$ at large
$x_F$.  In our view that mechanism is gluon
depletion.  Of course, if the $x_F$ dependence of $H(x_F,A)$ were
independently known, its incorporation in our analysis is straigtforward. For us here, we 
identify the $x_1$ dependence of $G(x_1, A)$ in Eq.(\ref{3}) with that
in Eq.(\ref{7}), which is completely known, and  proceed to
the study of the phenomenological implication on gluon depletion.

In the spirit of DGLAP evolution, even though the effect of  a nuclear target on the
projectile gluon distribution is highly non-perturbative, we now propose an evolution
equation on the gluon distribution $g (x,z)$, where $z$ is the path length in a nucleus.  For
the change of $g (x,z)$, as the gluon traverses a distance $dz$ in the nucleus, we write
\begin{eqnarray} 
{d  \over  dz}\,g(x,z) = \int^1_x {dx^{\prime}  \over 
x^{\prime}}\,g(x^{\prime},z)\,Q({x\over  x^{\prime}} ),
\label{8}
\end{eqnarray}
where $Q(x/x')$ describes the gain and loss of gluons in $dz$, but unlike the splitting
function in pQCD, it cannot be calculated in perturbation theory.  Equation (\ref{8}) is
similar to the nucleonic evolution equation proposed in \cite{hp}, except that this is now
at the parton level.  Instead of guessing the form of $Q(x/x')$, which is unknown, we
shall use  Eq.(\ref{7}) to determine it phenomenologically.

To that end, we first define the moments of $g(x,z)$ by
\begin{eqnarray} 
g_n(z) = \int_0^1dx\, x^{n-2}\, g(x,z).
\label{9}
\end{eqnarray} 
Taking the moments of  Eq.\ (\ref{8}) then yields
\begin{eqnarray} 
d g_n(z) / dz = g_n(z)\, Q_n,
\label{10}
\end{eqnarray} 
where $Q_n=\int_0^1 dy\,y^{n-2}\,Q(y)$.  It then follows that
\begin{eqnarray} 
g_n(z) = g_n(0)\,e^{zQ_n},
\label{11}
\end{eqnarray} 
whose exponential form suggests $Q_n<0$ for the physical process of depletion.  The gluon
depletion function $D(y,z)$ is defined by
\begin{eqnarray} 
g(x,z) = \int_x^1{dx'\over x'} g(x',0)\,D({x\over x'},z),
\label{12}
\end{eqnarray} 
where $g(x',0)$ is the gluon distribution in a free nucleon.  From Eq.\ (\ref{12}) we have
$g_n(z)=g_n(0)\,D_n(z)$, where $D_n(z)$ is the moment of $D(y,z)$.  Comparison with
Eq.\ (\ref{11}) gives
\begin{eqnarray} 
D_n(z) = e^{zQ_n}.
\label{13}
\end{eqnarray} 

To relate this result to $R(x_F,A)$, we first note that $G(x_F,A)$ in Eq.(\ref{3})
is, by definition, $G(x_F,A)=g(x_1,A)/g(x_1,0)$, where $x_F$ is expressed in terms of
$x_1$.  It then follows from Eq.(\ref{3})  that 
\begin{eqnarray}
J(x_1,A) &\equiv&
g(x_1,0)\,R(x_F(x_1),A) / N(x_2(x_1),A)\nonumber\\ &=& g(x_1,A)\,H(A).
\label{14}
\end{eqnarray} 
In relating $A$ to the average path length $L$ of the projectile $p$ through the nucleus,
we use $L=3R_A/2=1.8A^{1/3}$fm.  We then set $z=L/2$ for the average distance
traversed at the point of $c\bar c$ production.  Thus when referring to the last expression
of Eq.(\ref{14}), we write $J(x_1,A)=g(x_1,z(A))\,H(z(A))$, where $g(x_1,z)$ is to be
identified with that in Eq.(\ref{12}).  Note that the $A$ dependence of the middle
expression in Eq.(\ref{14}) is, on account of Eq.(\ref{7}), in terms of ln$A$, whereas that of
the last expression is in terms of $z$, or $A^{1/3}$.  Since it is known that ln$A\approx
A^{1/3}$ for $60<A<240$, we shall consider the consequences of Eq.(\ref{14}) only for $A$
in that range. We suggest that a revised form of presenting the data, different that in
Eq.(\ref{1}), should be tried in the future.

Taking the moments of $J(x_1,A)$, we get using Eq.(11)
\begin{eqnarray} 
{\rm ln}J_n(A) - {\rm ln}g_n(0) = zQ_n+{\rm ln}H(z).
\label{15}
\end{eqnarray} 
To determine $Q_n$, it is necessary to use as an input the gluon distribution $g(x_1,0)$ in
a free proton at $Q^2=10\, {\rm GeV}^2$.  We adopt the simple canonical form
\begin{eqnarray} 
g(x_1,0) = g_0(1-x_1)^5,
\label{16}
\end{eqnarray} 
where the constant $g_0$ is cancelled in Eq.(\ref{15}) due to the definition of
$J(x_1,A)$.  In our calculation we set $g_0=1$.  Indeed, the accuracy of $g(x_1,0)$ is
unimportant, since it enters Eqs.(\ref{14}) and (\ref{15}) in ways that render the result
insensitive to its precise form.  On the basis of Eqs.(\ref{7}) and (\ref{16}), $J(x_1,A)$ is
therefore known.  The LHS of Eq.(\ref{15}) can then be computed except for a caveat.  To
calculate the moments of $J(x_1,A)$, it is necessary to compute
$\int_0^1dx_1\,x_1^{n-2}\,J(x_1,A)$.  However, $x_1$ cannot be less than $\tau$ in order
to keep $x_2\leq 1$ [see Eq.(\ref{6})].  Furthermore, Eq.(\ref{7}) does not provide
reliable information on $J(x_1,A)$ at small $x_1$, since the parametrizations of
$\alpha(x_F)$ and $\beta(x_2)$ are for the variables in  ranges that exclude the
$x_1\rightarrow \tau$ limit.  Fortunately, that part of the integration in
$x_1$ can be suppressed by considering $n\geq 3$. 
The part of the integration in the interval $0<x_1<\tau$ amounts to only about 2\%
contribution even at $n=2$ (if naive extrapolation is used), so its inaccuracy will be
neglected.  Physically, it is the data at high $x_F$
that we emphasize in our analysis, and that corresponds to the high-$n$ moments of
$J(x_1,A)$. 

For convenience, let us denote the LHS of Eq.(\ref{15}) by $K_n(z)$, i.e., $K_n(z)\equiv
{\rm ln}[J_n(z(A))/g_n(0)]$.  For sample cases of $A=100\ {\rm and}\ 200$, they are shown
as discrete points in Fig.\ 1 for $3\leq n\leq 20$.  Instead of performing an inverse
Mellin transform on $K_n(z)$, our procedure is to fit $K_n(z)$ by a simple formula
that can yield
$Q(y)$ by inspection.  The fitted curves shown by the solid and dashed lines in Fig.\ 1
are obtained by use of the formula
\begin{eqnarray} 
K_n = -k_0+{k_1\over n}-{k_2\over n+1}+{k_3\over n+2}.
\label{17}
\end{eqnarray} 
Using $k_i$ and $k'_i$ to denote the values for the cases $A=100$ and 200, respectively,
we have
\begin{eqnarray} 
k_0=1.592, \quad k_1=23.42, \quad k_2=97.66,\quad k_3=89.17\nonumber\\
k'_0=1.831, \quad k'_1=27.43, \quad k'_2=113.97,\quad k'_3=103.80.\nonumber
\end{eqnarray} 
 Because of Eq.(\ref{15}), the $n$ dependence of $K_n$ prescribes the $n$
dependence of $Q_n$.  Let us therefore write
\begin{eqnarray} 
Q_n = -q_0+{q_1\over n}-{q_2\over n+1}+{q_3\over n+2}.
\label{18}
\end{eqnarray} 
Since Eq.(\ref{15}) is to be used only for $A>60$, we evaluate it at
$A=100$ and 200, and
take the difference.  Denoting $z$ by $z_1$ and $z_2$, respectively, for
the two $A$ values,
and with $\Delta k_i=k'_i-k_i,\ \Delta z=z_2-z_1$, we have
\begin{eqnarray}
\Delta k_0=q_0\,\Delta z-{\rm ln}{H(z_2)\over H(z_1)}, \quad  \Delta
k_i=q_i\,\Delta z,\ (i\neq 0).
\label{19}
\end{eqnarray}
For the hadron absorption factor $H(z)$ we write it in the canonical exponential form
\cite{gh}, $H(z)={\rm exp}(-\rho\sigma z)$, where $\rho^{-1}=(4/3)\pi(1.2)^3\,$fm$^3,
z=0.9A^{1/3}\,$fm, and $\sigma$ is the absorption cross section.  Putting these in
Eq.(\ref{19}), we get (with $\Delta z=1.086$ fm)
\begin{eqnarray} 
q_0+\rho\sigma=0.22,\ \  q_1=3.68,\ \ q_2=15.01,\ \ q_3=13.47
\label{20}
\end{eqnarray} 
 in units of fm$^{-1}$.
  
There is a reason why $q_0$ and $\rho\sigma$ enter Eq.(\ref{20}) as a sum.  To
appreciate the physics involved, we first note that Eq.(\ref{18}) implies directly
\begin{eqnarray} 
Q(y)=-q_0\,\delta(1-y)+q_1\,y-q_2\,y^2+q_3\,y^3.
\label{21}
\end{eqnarray} 
The first and third terms on the RHS above are the loss terms (i.e., gluon depletion),
while the second and last terms represent gain (i.e., gluon regeneration).  If $Q(y)$
consisted of only the first term, then using it in Eq.(\ref{8}) would give
$dg(x,z)/dz=-q_0\,g(x,z)$, whose solution is of the same exponential form as that of
absorption.  With both depletion and absorption present, the
exponents lead to a sum, as in Eq.(\ref{20}). Our $Q(y)$ is, however, more
complicated.  The $-q_2\,y^2$ term gives rise to depletion that depends on the shape
of $g(x,z)$, while the $q_1\,y+q_3\,y^3$ terms generate new gluons at $x$ from all the
gluons at $x'>x$. 

 Since $Q_n$ decreases
monotonically with $n$, we require $Q_3<0$, and exclude $Q_2$ from this
consideration because of its inaccuracy discussed earlier. Combining Eqs.(18) and
(20), we get $\rho\sigma<q_0+\rho\sigma-q_1/3+q_2/4-q_3/5=0.05$ fm$^{-1}$.  We
thus set $q_0=0.17$ fm$^{-1}$.

Since it is not easy to see directly from $Q_n$ or $Q(y)$ the magnitude of the effect of
gluon depletion and regeneration, we can calculate $g(x_1,z)$, not from Eq.(\ref{12}),
but by fitting the calculated
$g_n(z)$ in Eq.(\ref{11}), using the formula
\begin{eqnarray}
g_n(z)=\sum_{i=1}^3 a_i(z)\ B(n-1,5+i),
\label{22}
\end{eqnarray}  
 where $B(a,b)$ is the beta function.  Then the result yields directly
\begin{eqnarray}
g(x_1,z)=\sum_{i=1}^3\ a_i(z)\ (1-x_1)^{4+i}.
\label{23}
\end{eqnarray}
For $A=100\ (200)$, i.e., $z=z_1\ (z_2)$, we have
$a_1=0.58\ (0.485), a_2=0.92\ (1.118)$, and $a_3=-0.47\ (-0.56)$ for $g_0=1$ in
Eq.(\ref{16}). The result for
$G(x_1,z)=g(x_1,z)/g(x_1,0)$ is then 
\begin{eqnarray}
G(x_1,z)=a_1(z)+a_2(z)(1-x_1)+a_3(z)(1-x_1)^2,
\label{24}
\end{eqnarray}
which is  shown in Fig.\ 2 for two values of $A$.  It is
now evident that gluon depletion suppresses the gluon distribution at medium and high
$x_1$, but the unavoidable gluon regeneration enhances the distribution at low $x_1$. 
The cross-over occurs at $x_1\approx 0.28$.

Let us now exhibit our result for $\alpha(x_F)$, which is shown in
Fig.3.  The line is obtained by use of 
Eq.(\ref{24}) in Eq.(\ref{3}) and $\sigma=6.5$ mb in $H(z)$. Only one line is shown
for both $A=100$ and 200, their difference being negligible in the plot. 
Since our method of using the moments cannot be extended to
$n=2$ due to the problems mentioned after Eq.(\ref{16}), there is some inaccuracy
inherent in our analysis. Thus the fit cannot be expected to be perfect. 
Our model can reproduce the general trend of the $x_F$ dependence, but not the
detail structure, for which more terms in Eqs.(\ref{17}) and (\ref{18}) would be
needed.   The overall suppression is achieved by use of a phenomenological value of
$\sigma$, rather than the bound based on the technical  assumption of $Q_3<0$. 

Our analysis has been based on the assumption that $H(z,A)$ is independent
of $x_F$.  If and when that $x_F$ dependence can be determined independently,
the effect can easily be incorporated in our analysis to modify our numerical result . 
Since that dependence is not likely to be strong \cite{hhk,a,v2}, the modification
would be minor.  Our study shows that the $J/\psi$ suppression observed at large
$x_F$ in
$pA$ collisions \cite{l} strongly suggests the presence of gluon depletion in the beam
proton at high $x_1$.  The significance of this finding goes beyond the $J/\psi$
suppression problem itself, since it would revise the conventional thinking concerning
the role of partons in nuclear collisions. 
 
Since the gluon distribution is enhanced for $x_1\leq 0.28$, the $J/\psi$ suppression
observed in the $x_F\approx  0$ region  in the heavy-ion collisions at CERN-SPS
cannot be due to the gluon depletion effect.  The same would be true at RHIC. 
However, we expect a significant increase in suppression at large $x_F$ due to gluon
depletion, not to color deconfinement.  We further speculate at this point that the
gluon enhancement at low $x$ may be responsible, at least in part, for the
strangeness and dilepton enhancement already observed in heavy-ion collisions.

This work was supported, in part, by the U.S.-Slovakia Science and Technology Program,
National Science Foundation under Grant No. INT-9319091 and by the U. S. Department of
Energy under Grant No. DE-FG03-96ER40972.

\begin{figure}
\caption { $K_n$. The curves are fitted results using Eq.(17).}
\end{figure}

\begin{figure}
\caption {$G(x_1,z)$ showing the effects of gluon depletion.}
\end{figure}

\begin{figure}
\caption {$\alpha(x_F)$ vs $x_F$. The solid line is our result compared to the data
from [3].}
\end{figure}

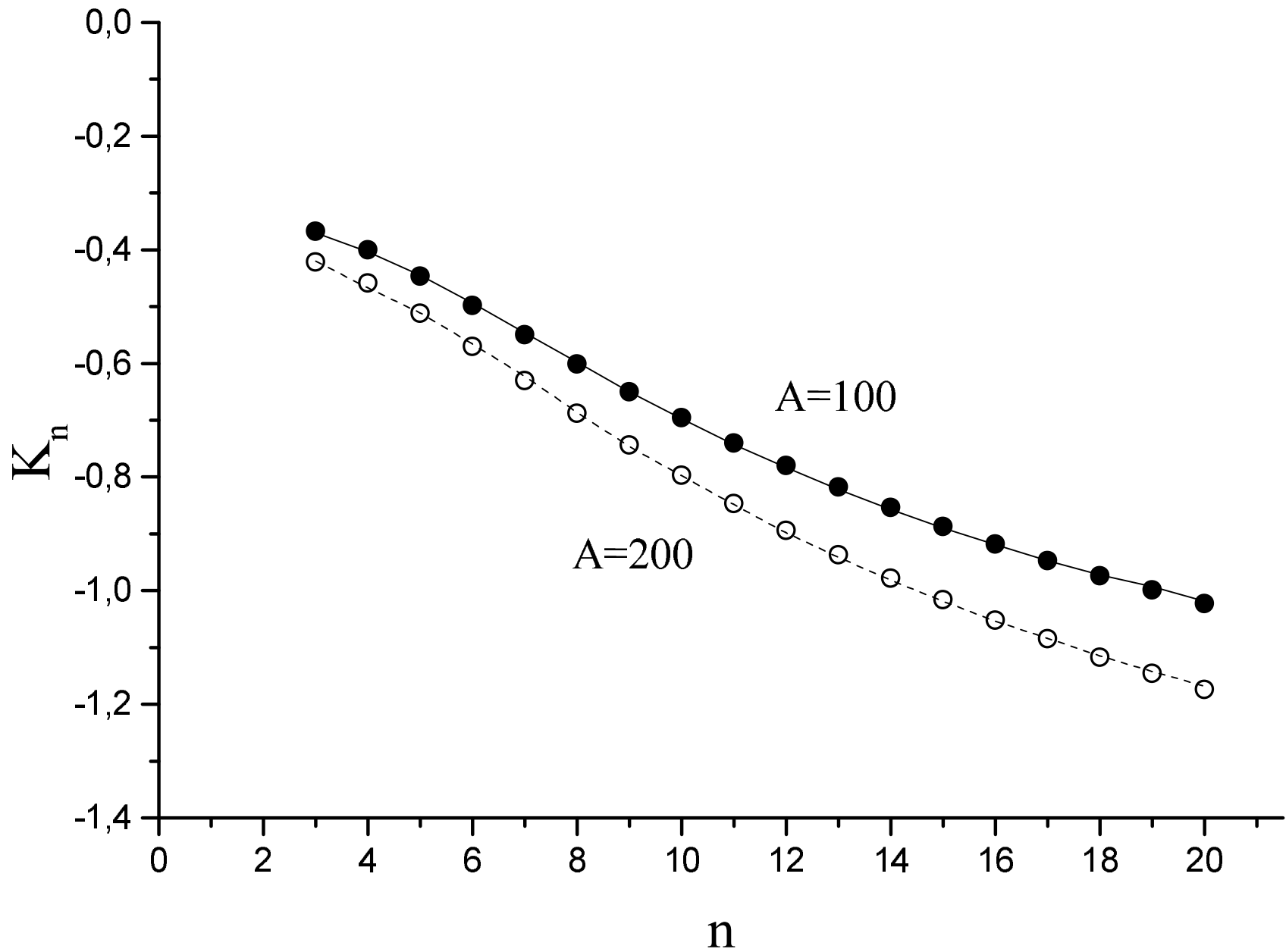
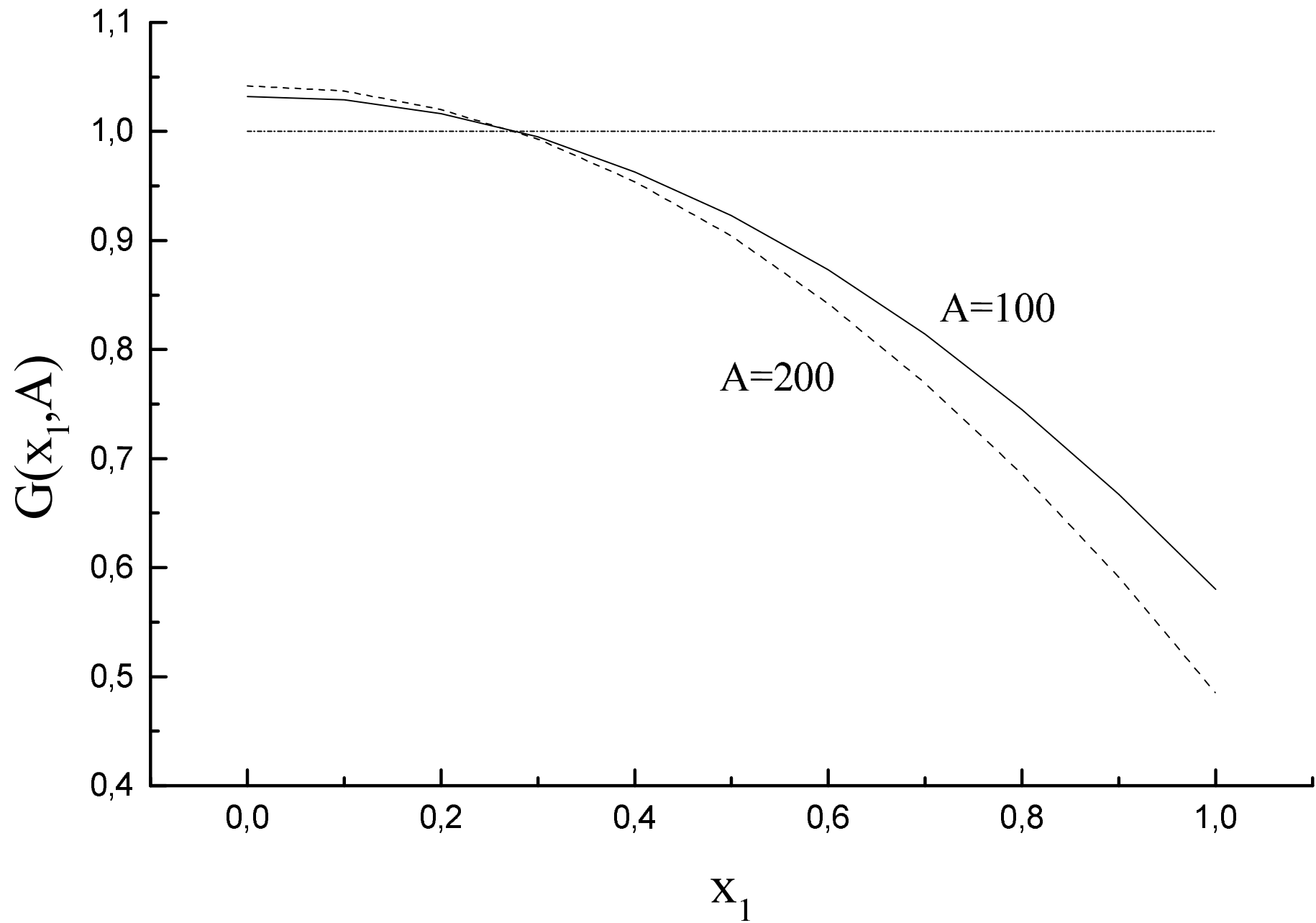
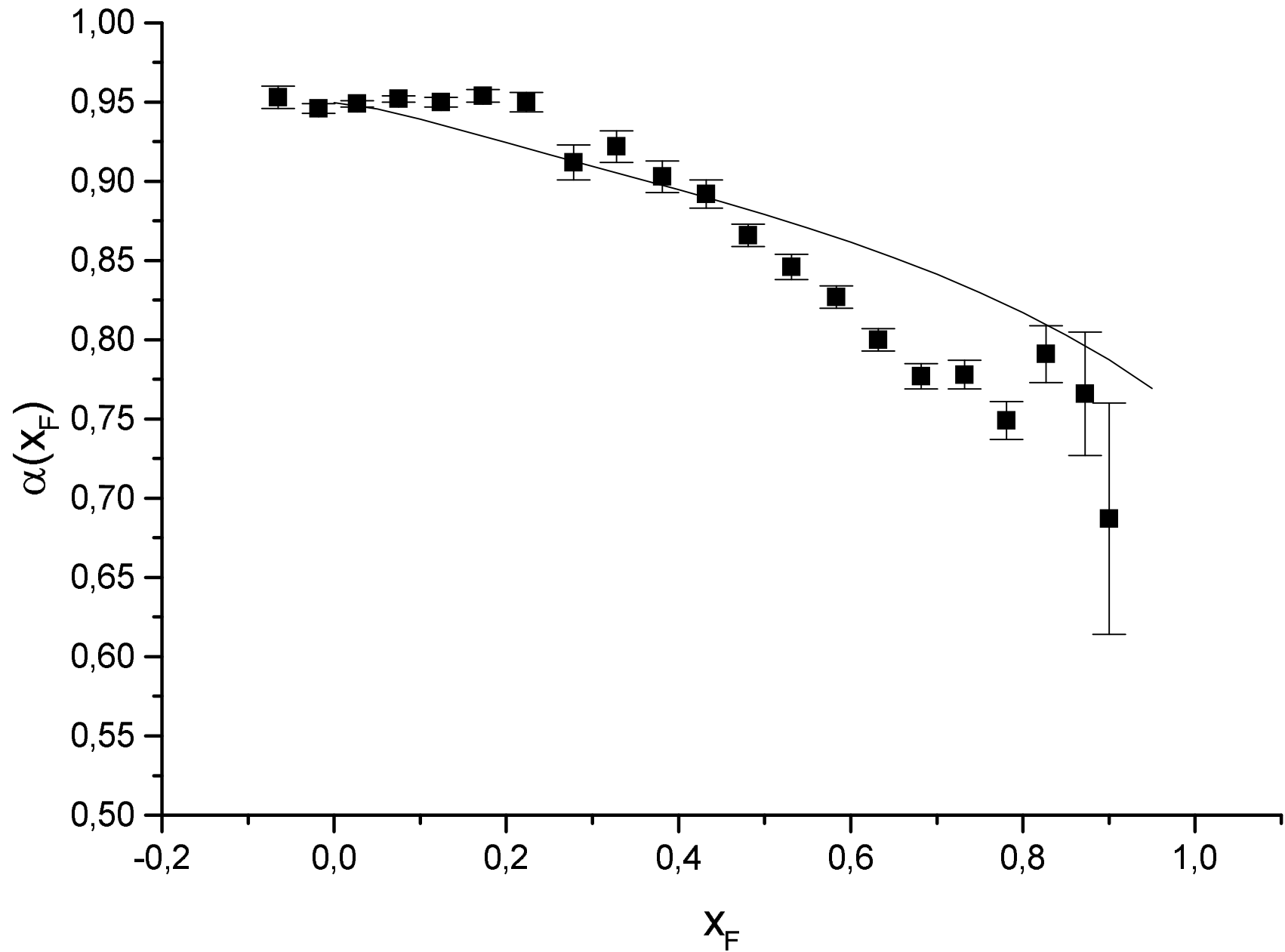
\end{document}